\documentclass[12pt,preprint]{aastex}

\usepackage{verbatim} 

\slugcomment{}

\shorttitle{H$\alpha$ Luminosity Class Diagnostic}
\shortauthors{Jennings \& Levesque}

\begin{document}

\title{H$\alpha$ as a Luminosity Class Diagnostic for K- and M-type Stars}

\author{Jeff Jennings\altaffilmark{1} and Emily M. Levesque\altaffilmark{2}}

\begin{abstract}
We have identified the H$\alpha$ absorption feature as a new spectroscopic diagnostic of luminosity class in K- and M-type stars. From high-resolution spectra of 19 stars with well-determined physical properties (including effective temperatures and stellar radii), we measured equivalent widths for H$\alpha$ and the Ca II triplet and examined their dependence on both luminosity class and stellar radius. H$\alpha$ shows a strong relation with both luminosity class and radius that extends down to late M spectral types. This behavior in H$\alpha$ has been predicted as a result of the density-dependent overpopulation of the metastable 2$s$ level in hydrogen, an effect that should become dominant for Balmer line formation in non-LTE conditions. We conclude that this new metallicity-insensitive diagnostic of luminosity class in cool stars could serve as an effective means of discerning between populations such as Milky Way giants and supergiant members of background galaxies.
\end{abstract}

\section{Introduction}
\footnotetext[1]{CASA, Department of Astrophysical and Planetary Sciences, University of Colorado, 389-UCB, Boulder, CO 80309, USA}
\footnotetext[2]{Department of Astronomy, University of Washington, Seattle, WA 98195, USA \\Corresponding author: \texttt{emsque@uw.edu}}

Luminosity class diagnostics are critically important for the study of cool (K- and M-type) stars both in and beyond the Milky Way. An effective luminosity class discriminant is necessary in order to distinguish between foreground dwarf and giant populations and background supergiants, particularly when studying massive star populations in neighboring Local Group galaxies. Massey (1998) illustrated this, determining that $\sim$50\% of the red stars included in the Humphreys \& Sandage (1980) survey of M33 were actually foreground dwarfs rather than red supergiant members of M33. In this case, the foreground dwarf and background supergiant populations were separated by the use of a ($B-V$) vs. ($V-R$) color-color diagram, which revealed a clear separation in ($B-V$) as a consequence of surface-gravity-dependent line blanketing effects. However, while this color-based method can be useful for separating stars at extreme ends of the luminosity class regime, delving into a more precise distinction between dwarf, giant, and supergiant requires the increased precision of spectroscopic diagnostics.

Absorption features of neutral and singly-ionized metal lines are commonly cited as luminosity class diagnostics in cool stars (e.g. Keenan \& Hynek 1945, Rayner et al. 2009, Negueruela et al.\ 2012). These include the strong Ca I, Mg I, and Na D absorption features (e.g. Smith \& Drake 1987, Fuhrmann et al.\ 1997), but the most well-studied is the Ca II triplet (8498, 8542, 8662\AA). This absorption feature has been shown to be strongly sensitive to surface gravity, with the three absorption lines all getting stronger at lower surface gravities (see Cenarro et al.\ 2001a,b and references therein). This relation has been used to identify supergiants both in the Milky Way and in Local Group galaxies (e.g. Garzon et al.\ 1997; Mantegazza 1992; Massey 1998; Rayner et al.\ 2009; Negueruela et al.\ 2012; Britavskiy et al.\ 2014, 2015).

Unfortunately there are several drawbacks to using the Ca II triplet as a means of distinguishing between dwarfs, giants, and supergiants, particularly across multiple host galaxies. The Ca II triplet feature is sensitive to metallicity (e.g. Armandroff \& Da Costa 1991, Battaglia et al.\ 2008, Starkenburg et al.\ 2010, Sakari \& Wallerstein 2016). While this effect is small compared to luminosity class effects between dwarfs and supergiants (e.g. Mallik 1997, Massey 1998), it can lead to difficulties distinguishing between intermediate classes, such as dwarfs and giants or giants and supergiants. As an example, Massey (1998) noted that low-metallicity supergiants and solar-metallicity giants have comparable equivalent widths. This ambiguity introduces the possibility of confusing foreground Milky Way red giants with lower-metallicity background red supergiants, a common scenario when trying to separate halo giants from background supergiant populations of metal-poor Local Group galaxies such as the Magellanic Clouds, NGC 6822, and WLM (e.g. Levesque et al.\ 2006, 2007; Levesque \& Massey 2012; Britavskiy et al.\ 2014, 2015). In addition, the Ca II triplet's sensitivity to luminosity class has not been calibrated for spectral types later than M4 (e.g. Negueruela et al.\ 2012), and observations show evidence that this relation may break down beyond early M spectral types (e.g. Rayner et al.\ 2009). Finally, the Ca II triplet lines can be contaminated by the hydrogen Paschen lines; while this contamination is corrected for in the Cenarro et al.\ (2001a) calibration of the Ca II triplet index with luminosity class, in practice it is often difficult to distinguish between these features in observations (e.g. Britavskiy et al.\ 2014).

Here we present an examination of Ca II triplet and H$\alpha$ absorption features in a sample of 19 K- and M-type dwarfs, giants, and supergiants. We drew our sample from stars with pre-determined effective temperatures and stellar radii and acquired new high-resolution ($R \sim 31500$) spectra (Section 2). From these spectra and the stars' previously-determined physical properties we were able to measure the equivalent widths of both features and examine their dependence on both luminosity class and radius. In addition to confirming previous work on the Ca II triplet, we also find that the H$\alpha$ absorption feature serves as an excellent diagnostic of both luminosity class and radius in stars extending down to late M spectral types (Section 3), a result of density-dependent overpopulation of the metastable 2$s$ level in hydrogen. We consider the results' implications and potential applications for future work, particularly in discerning between foreground giant and background supergiant populations (Section 4).

\section{Observations}
\subsection{Sample Selection and Observations}
Our sample has been drawn principally from van Belle et al. (1999, 2009a, 2009b), who  presented interferometrically-determined radii and effective temperatures ($T_{\rm eff}$) for cool dwarf (class V), giant (III), and supergiant (I) stars. These ``direct" radii determinations combined with the well-known nature of the samples (i.e., precise metallicities for the dwarfs and giants where metallicity evolution may be substantial and distances to host OB associations for the supergiants) make them an ideal resource for testing luminosity class diagnostics. From these studies, along with three additional cool stars with precisely determined $T_{\rm eff}$ values from Cenarro et al. (2001a), we have a sample of 19 stars.

To avoid confusing luminosity class effects with $T_{\rm eff}$-dependent changes in line strengths, we split our sample into four discrete bins according to previously-determined $T_{\rm eff}$: 3580 K, 3650 K, 4050 K, and 4150 K. Bin identifiers are the mean $T_{\rm eff}$ of each bin rounded to the nearest 10 K; stars within each bin span a total range of $\le$50 K. Changes in spectral features with luminosity class were then compared within each bin. Our full sample, sorted by bins, is given in Table 1.

We observed our sample of cool stars using the Astrophysics Research Consortium Echelle Spectrograph (ARCES; Wang et al.\ 2003) on the Apache Point Observatory 3.5-meter telescope, using the default $1\arcsec.6 \times 3\arcsec.2$ slit for an $R\sim31500$. On 2014 July 29 we observed 11 targets under clear conditions with a seeing of $1\arcsec.1$; on 2015 February 5 we observed 8 targets under clear conditions with a seeing of $1\arcsec.0$. To achieve precise flatfield and wavelength calibrations for each star, we observed quartz lamps and ThAr lamps, as well as telluric standards. The spectra were reduced using standard IRAF{\footnote{IRAF is distributed by NOAO, which is operated by AURA, Inc., under cooperative agreement with the NSF.}} procedures, using the ThAr lamp spectra for wavelength calibrations.

\section{Equivalent Widths and Analyses}
We measured the equivalent widths of the Ca II triplet (summing the three lines, hereafter $W$(CaT); Mallik 1997, Cenarro et al.\ 2001a) and the H$\alpha$ line (hereafter $W$(H$\alpha$)) using IRAF. We found that Voigt profiles yielded the best fits to the absorption features (see also Sakari \& Wallerstein 2016 for a discussion of Ca II triplet line profiles), with a systematic measurement error of $\sim$1\%; our $W$(CaT) values show good agreement with those measured by Mallik (1997) and Massey (1998). Table 1 includes measured $W$(CaT)and $W$(H$\alpha$) values for each star in our sample.

The star HD 207991 was originally included in our sample as a supergiant star of luminosity class I, based on the spectral type reported in van Belle et al.\ (2009b); however, both the radius measured in van Belle al.\ (2009b) and our measured equivalent widths are more consistent with the star being a class III rather than a class I. Van Belle et al.\ (2009b) specifically singles out this star as being unusually small for a supergiant and suggests that it may be misclassified. Furthermore, the original reference for the spectral type of HD 207991, Keenan \& McNeil (1989), classifies it as a K4 III star. Based on this we classify HD 207991 here as a giant with luminosity class III.  

\subsection{The Dependence of H$\alpha$ On Luminosity Class}
 
While the dependence of $W$(CaT) on luminosity has been previously studied in cool stars (e.g. Cenarro et al. 2001a), this is the first study of such behavior in $W$(H$\alpha$). The effect is a predicted consequence of the density-dependent overpopulation of the metastable 2$s$ level in hydrogen, first described in detail in Struve et al.\ (1939) and further discussed in Huang et al.\ (2012). Struve et al.\ (1939) posit that observations of strong Balmer absorption lines in late type supergiants may be explained by the metastability of hydrogen's 2$s$ state, a consequence of the classically forbidden $2s \rightarrow 1s$ radiative transition. They suggest that measurements of the Paschen lines may confirm this, with the expectation that the widths of these lines, originating from the $n = 3$ level (i.e., states that are not metastable), would not show the broadening observed for H$\alpha$. Huang et al.\ (2012) measure the equivalent widths of  Paschen delta (P$\delta$) and H$\alpha$ for G-type and later stars, observing P$\delta$ widths consistent with model spectra that assume LTE. By contrast, they observe substantially broader H$\alpha$ widths than predicted under LTE and note that because of the forbidden $2s \rightarrow 1s$ transition, the population of the metastable 2$s$ level in non-LTE is largely determined by radiative and collisional transitions with other excited states. In low density conditions where the mean free path is larger (such as the atmospheres of cool giant and supergiant stars with lower surface gravities) such transitions are less frequent, leading to an overpopulation of the metastable 2$s$ state. This effect in cool, late type stars should then scale inversely with surface gravity and, consequently, luminosity class, resulting in an H$\alpha$ line width that increases as we move from class V giants to class I supergiants. This is consistent with our observations, as shown in Figure 1.

The evolution of $W$(H$\alpha$) is consistent across our $\mathrm{T_{eff}}$ bins, supporting the notion that this overpopulation effect dominates temperature effects for both K- and M-stars (in agreement with Struve et al. 1939 and Huang et al. 2012). A connection between luminosity class and the strength of the H$\alpha$ line has also been observed in narrow-band survey photometry (e.g. Drew et al.\ 2005, 2014; Wright et al.\ 2008).

\subsection{A Comparison of Luminosity Class Diagnostics}
Figure 2 compares $W$(CaT) to luminosity class (top) and stellar radius (bottom) for stars in each of our four $T_{\rm eff}$ bins. As seen in previous work, $W$(CaT) shows a clear relation with luminosity class, with a mean fractional difference between classes I and V across $T_{\rm eff}$ bins of $-$0.80. We do, however, see that this relation weakens at lower $T_{\rm eff}$ (in the 3580 K bin the piecewise relation shows much better agreement than the relatively poor linear fit as a consequence of the inability to clearly distinguish between giants and supergiants). This is in agreement with the results of Rayner et al.\ (2009) and Negueruela et al.\ (2012) suggesting that the $W$(CaT)-luminosity class relation breaks down beyond early M spectral types (3580 K corresponds to a spectral type of M3-3.5 in supergiants; Levesque et al.\ 2005).

Figure 3 illustrates the same comparisons to luminosity class and stellar radius for $W$(H$\alpha$). We see that the $W$(H$\alpha$) feature shows a comparable relation with luminosity class, yielding a mean fractional difference between classes I and V of $-$1.00. This persists in all four $T_{\rm eff}$ bins, ranging from $-$0.83 in the 4050 K bin to $-$1.43 in the 3580 K bin.

The relation between these spectral features and stellar radius is harder to quantify; however, we expect the behavior of $W$(CaT) and $W$(H$\alpha$) to be inversely dependent on surface gravity, with $W \propto R^2$. The small sample sizes preclude any possibility of a robust polynomial fit within any single bin. We instead normalize the data points in each bin and combine the full dataset to determine best-fit second-order polynomials for both the $W$(CaT) and $W$(H$\alpha$) relations. As illustrated in Figure 4, we find that while a $\sim R^2$ relation offers the best fit for both datasets (as opposed to linear, exponential, or higher-order fits), the fit is better for the $W$(H$\alpha$) data with a lower residual standard deviation (17\% for $W$(H$\alpha$) as opposed to 25\% for $W$(CaT)).

\section{Discussion}
From our analyses of high-resolution spectra for 19 K- and M-type stars, we have identified $W$(H$\alpha$) as a new diagnostic of luminosity class and radius in cool stars. This agrees with predictions stating the density-dependent overpopulation of the metastable 2$s$ level hydrogen should become a dominant effect in non-LTE conditions, leading to a broadening of H$\alpha$ at lower surface gravities (Struve et al.\ 1939, Huang et al.\ 2012).

Reliable spectroscopic diagnostics of luminosity class for cool stars are particularly valuable in the current era of extragalactic stellar spectroscopy. Contamination of extragalactic red supergiant samples by foreground dwarfs can be mitigated through the use of color-color diagrams (Massey 1998); however, distinguishing between background supergiants and foreground giants can only be done through spectroscopic analyses. The utility of $W$(CaT) in discerning between giants and supergiants is limited. Differences in metallicity between the foreground and background populations can complicate interpretation of the Ca II line widths, and contamination by the nearby Paschen lines in both absorption and, in some cases, emission (see, for example, Castelaz et al.\ 2000, Levesque et al.\ 2014) makes the equivalent widths harder to measure. At the coolest temperatures we also find that the $W$(CaT) relation with luminosity class becomes less pronounced when comparing giants and supergiants.

By comparison, $W$(H$\alpha$) is a good diagnostic of luminosity class across the full range of temperatures represented in our sample and should be independent of metallicity effects. The robust relation between $W$(H$\alpha$) and stellar radius suggests this feature does indeed scale as expected for a direct tracer of luminosity and that it is a more effective tracer than $W$(CaT). While H$\alpha$ is certainly a weaker spectral feature than the Ca II triplet for K- and M-type stars, high-resolution observations of either feature are still required for use as luminosity diagnostics; acquiring sufficiently high-S/N and high-resolution observations of H$\alpha$ is attainable for populations extending out to extragalactic supergiants in the Local Group.

The applications of $W$(H$\alpha$) as a luminosity class diagnostic may also extend to earlier spectral types; Huang et al.\ (2012) note that the non-LTE effects observed here in H$\alpha$ are also expected in G-type stars. Extending this analysis to G and possibly even F-type stars would be useful for identifying the temperature limits of this phenomenon and could be extremely valuable for the study of extragalactic yellow supergiant populations. For these warmer stars separating background class I and foreground class III and V stars is particularly challenging due to a lack of color-color diagnostics and Paschen line contamination of the Ca II triplet (see Neugent et al.\ 2010, 2012); the strong Balmer absorption features in these populations could offer a compelling alternative diagnostic. Finally, future work studying $W$(CaT) and $W$(H$\alpha$) for a larger sample of well-studied Galactic K- and M-type stars will allow us to quantify the relation between these spectroscopic features and stellar properties such as mass, metallicity, radius, and surface gravity to much greater precision.

This paper is based on data gathered with the Apache Point Observatory
3.5-meter telescope, which is owned and operated by the
Astrophysical Research Consortium. We gratefully acknowledge the unparalleled expertise of George Wallerstein shared during conversations related to this work, as well as useful discussions with Adam Ginsburg, David Pitman, and Allison Youngblood. Helpful comments by the referee,
Philip Massey, have improved the presentation of our arguments in several places. JJ thanks C. Davis for her guidance. JJ was supported in part by an Undergraduate Research Opportunities Program Research Assistantship from the University of Colorado. EML was supported in part by NASA through Hubble Fellowship grant number HST-HF-51324.01-A from the Space Telescope Science Institute, which is operated by the Association of Universities for Research in Astronomy, Incorporated, under NASA.

\begin{deluxetable}{l c c c c c c c c c}
\tabletypesize{\scriptsize}
\tablewidth{0pc}
\tablenum{1}
\tablecolumns{8}
\tablecaption{\label{tab:stars} Program Stars and Equivalent Widths}
\tablehead{
\colhead{Star}
&\colhead{Luminosity}
&\colhead{$T_{\rm eff}$}
&\colhead{Radius}
&\multicolumn{5}{c}{\textit{W} (\AA)}
&\colhead{Refs\tablenotemark{a}} \\ \cline{5-9}
\colhead{}
&\colhead{class}
&\colhead{(K)}
&\colhead{($R_{\odot}$)}
&\colhead{8498 \AA}
&\colhead{8542 \AA}
&\colhead{8662 \AA}
&\colhead{Total}
&\colhead{H$\alpha$}
&\colhead{}
}
\startdata
       {\bf 3580 K} & & & & & & & \\ \hline
        HD 236979 & I & 3574 &524 $\pm$ 175 & 1.87 $\pm$ 0.08 & 3.92 $\pm$ 0.10 & 3.21 $\pm$ 0.07 & 9.01 $\pm$ 0.15 & 1.32 $\pm$ 0.09 & 4\\
        IRC +30465 & III & 3595 &69.9 $\pm$ 14.5 & 1.62 $\pm$ 0.05 & 3.10 $\pm$ 0.06 & 2.83 $\pm$ 0.05 & 7.56 $\pm$ 0.09 & 0.95 $\pm$ 0.10 & 1\\
        IRC +30468 & III & 3585 &60.5 $\pm$ 12.8 & 1.86 $\pm$ 0.04 & 3.87 $\pm$ 0.06 & 3.13 $\pm$ 0.05 & 8.87 $\pm$ 0.09 & 1.00 $\pm$ 0.09 & 1\\
        BD+44 2051 & V & 3545 &\nodata & 0.57 $\pm$ 0.03 & 1.47 $\pm$ 0.03 & 1.10 $\pm$ 0.03 & 3.25 $\pm$ 0.06 & 0.35 $\pm$ 0.05 & 2\\
        HD 95735 & V &  3593 &0.395 $\pm$ 0.013 & 0.62 $\pm$ 0.04 & 1.34 $\pm$ 0.03 & 1.30 $\pm$ 0.03 & 3.14 $\pm$ 0.06 & 0.36 $\pm$ 0.08 & 3 \\ 
        {\bf 3650 K} & & & & & & & \\ \hline
        HD 14404 & I & 3647 &405 $\pm$ 137 & 2.39 $\pm$ 0.08 & 4.68 $\pm$ 0.07 & 3.82 $\pm$ 0.07 & 10.88 $\pm$ 0.12 & 1.69 $\pm$ 0.11 & 4\\
        {HD 52005} & {I} & 3619 &266 $\pm$ 82 & 2.07 $\pm$  0.08 & 4.91 $\pm$  0.08 & 3.54 $\pm$  0.07 & 10.52 $\pm$ 0.13 & 1.54 $\pm$  0.09 & 4\\
        HR 6227 & III & 3640 &\nodata & 1.67 $\pm$ 0.05 & 3.58 $\pm$ 0.06 & 2.95 $\pm$ 0.05 & 8.16 $\pm$ 0.10 & 1.07 $\pm$ 0.10 & 2\\
        IRC +40022 & III & 3629 &\nodata & 1.72 $\pm$ 0.06 & 3.63 $\pm$ 0.06 & 2.82 $\pm$ 0.05 & 8.20 $\pm$ 0.09 & 1.02 $\pm$ 0.10 & 1\\
        HD 119850 & V &  3664 &0.481 $\pm$ 0.040 & 0.78 $\pm$ 0.03 & 1.65 $\pm$ 0.03 & 1.19 $\pm$ 0.04 & 3.61 $\pm$ 0.06 & 0.50 $\pm$ 0.07 & 3\\
        {\bf 4050 K} & & & & & & & \\ \hline   
       {HD 13686} & {I} & 4054 &99 $\pm$ 34 & 2.25 $\pm$ 0.07 & 4.87 $\pm$ 0.07 & 3.84 $\pm$ 0.05 & 10.97 $\pm$ 0.11 & 1.47 $\pm$ 0.09 & 4\\
       {HD 207991} & {III\tablenotemark{b}} & 4035 &39 $\pm$ 6 & 1.57 $\pm$ 0.05 & 3.60 $\pm$ 0.05 & 2.77 $\pm$ 0.04 & 7.95 $\pm$ 0.08 & 1.19 $\pm$ 0.07 & 4\\
        {HR 3249} & {III} & 4037 &\nodata & 1.67 $\pm$ 0.06 & 4.00 $\pm$ 0.06 & 2.87 $\pm$ 0.05 & 8.54 $\pm$ 0.10 & 1.21 $\pm$ 0.07 & 2\\
        {HR 7237} & {III} & 4075 &45.2 $\pm$ 5.1 & 1.78 $\pm$ 0.06 & 3.75 $\pm$ 0.05 & 2.92 $\pm$ 0.05 & 8.45 $\pm$ 0.09 & 1.14 $\pm$ 0.09 & 1\\
        {HD 157881} & {V} &  4030 &0.564 $\pm$ 0.068 & 1.22 $\pm$ 0.04 & 2.71 $\pm$ 0.04 & 2.22 $\pm$ 0.05 & 6.15 $\pm$ 0.07 & 0.61 $\pm$ 0.06 & 3\\
        {\bf 4150 K} & & & & & & & \\ \hline        
        {HD 207119} & {I} & 4154 &235 $\pm$ 92 & 2.12 $\pm$ 0.06 & 5.16 $\pm$ 0.07 & 3.75 $\pm$ 0.07 & 11.03 $\pm$ 0.11 & 1.53 $\pm$ 0.10 & 4\\
        {HR 6258} & {III} & 4134 &71.3 $\pm$ 14.4 & 1.96 $\pm$ 0.05 & 4.02 $\pm$ 0.06 & 3.23 $\pm$ 0.05 & 9.21 $\pm$ 0.10 & 1.31 $\pm$ 0.10 & 1\\
        {HR 389} & {III} & 4144 &25.1 $\pm$ 2.2 & 1.54 $\pm$ 0.07 & 3.26 $\pm$ 0.06 & 2.72 $\pm$ 0.07 & 7.52 $\pm$ 0.12 & 1.16 $\pm$ 0.09 & 1\\
        {HD 88230} & {V} & 4156 &0.649 $\pm$ 0.028 & 1.11 $\pm$ 0.04 & 2.36 $\pm$ 0.03 & 2.08 $\pm$ 0.04 & 5.55 $\pm$ 0.06 & 0.62 $\pm$ 0.04 & 3\\
\enddata
\tablenotetext{a}{$T_{\rm eff}$ and stellar radii (where applicable) for these stars are taken from from (1) van Belle et al.\ (1999), (2) Cenarro et al.\ (2001a), (3) van Belle et al.\ (2009a), and (4) van Belle et al.\ (2009b).}
\tablenotetext{b}{The luminosity class of this star is taken from Keenan \& McNeil (1989); for more discussion see Section 3.}
\end{deluxetable}

\begin{figure}
\epsscale{1}
\plotone{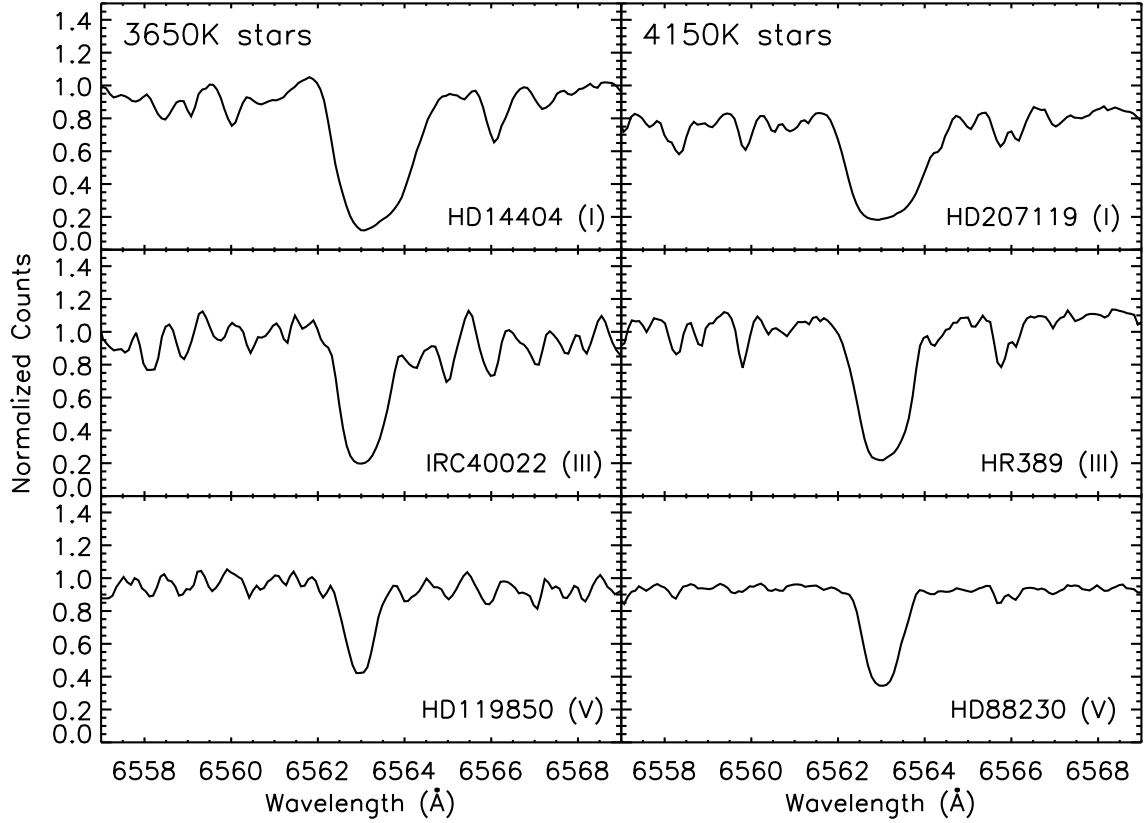}
\caption{Evolution of the H$\alpha$ absorption feature in a representative sample of our stars. In both the 3650 K bin (left) and 4150 K bin (right), we see $W$(H$\alpha$) decrease with increasing luminosity class (larger radius, lower surface gravity).}
\end{figure}

\begin{figure}
\epsscale{1}
\plotone{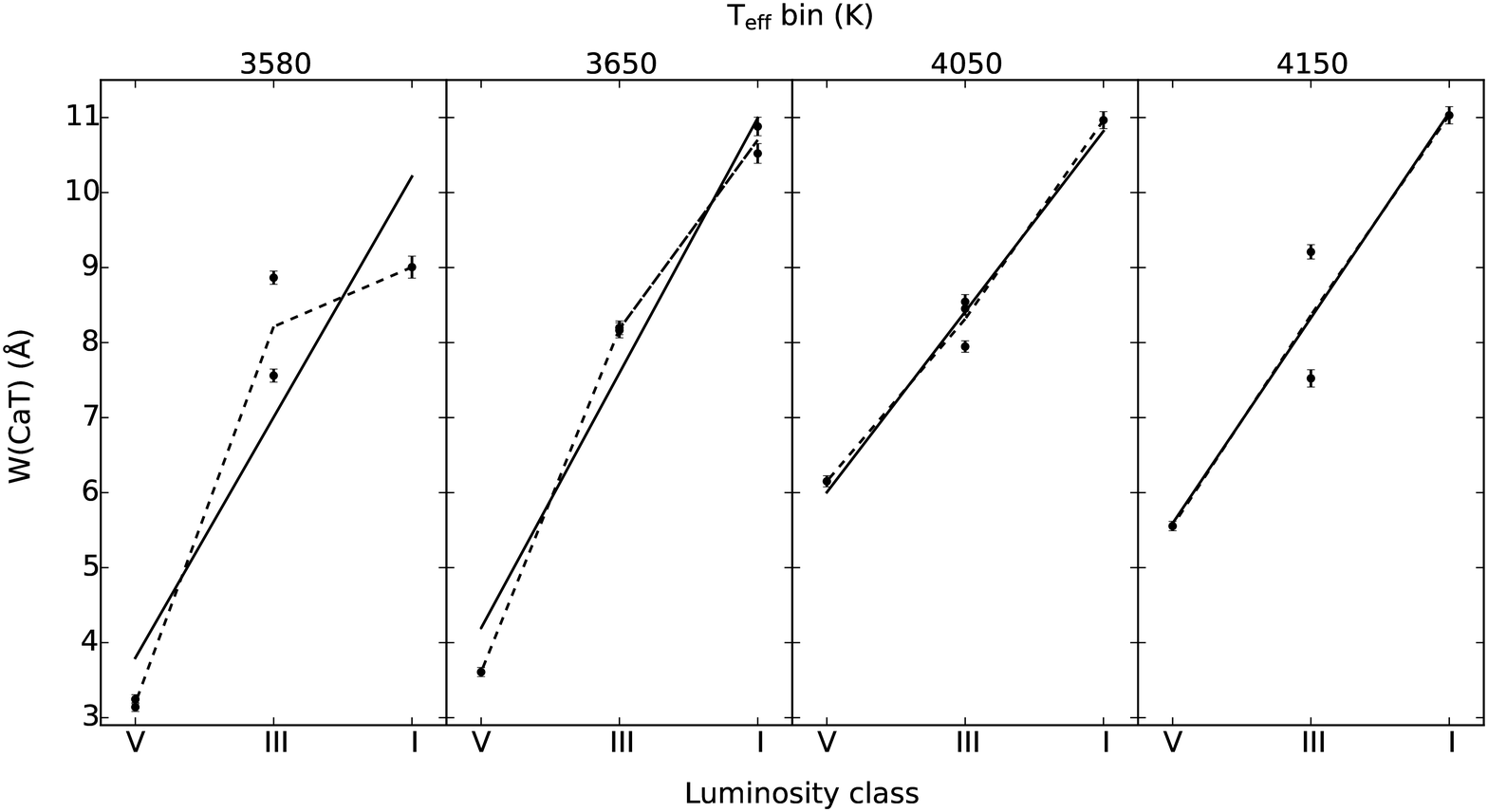}
\plotone{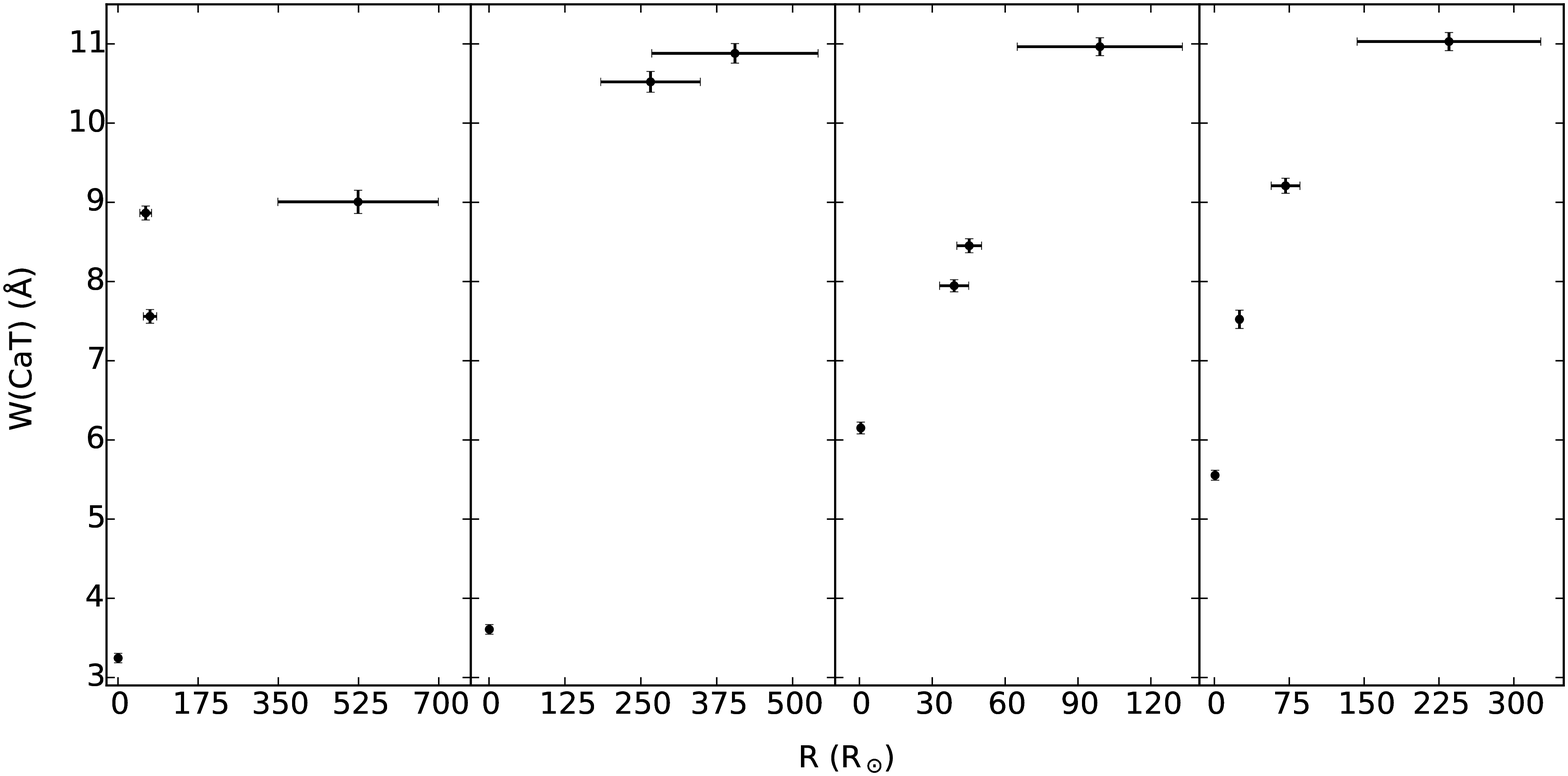}
\caption{\textit{W}(CaT), the sum of the Ca II triplet equivalent widths, for each star in our sample versus luminosity class (top) and stellar radius where available (bottom). Solid lines are linear fits to all stars in a given effective temperature bin, dashed lines are piecewise fits. The mean fractional difference in \textit{W}(CaT) is $-$0.80 between luminosity classes I and V, $-$0.24 between I and III, and $-$0.56 between III and V.}
\end{figure}

\begin{figure}
\epsscale{1}
\plotone{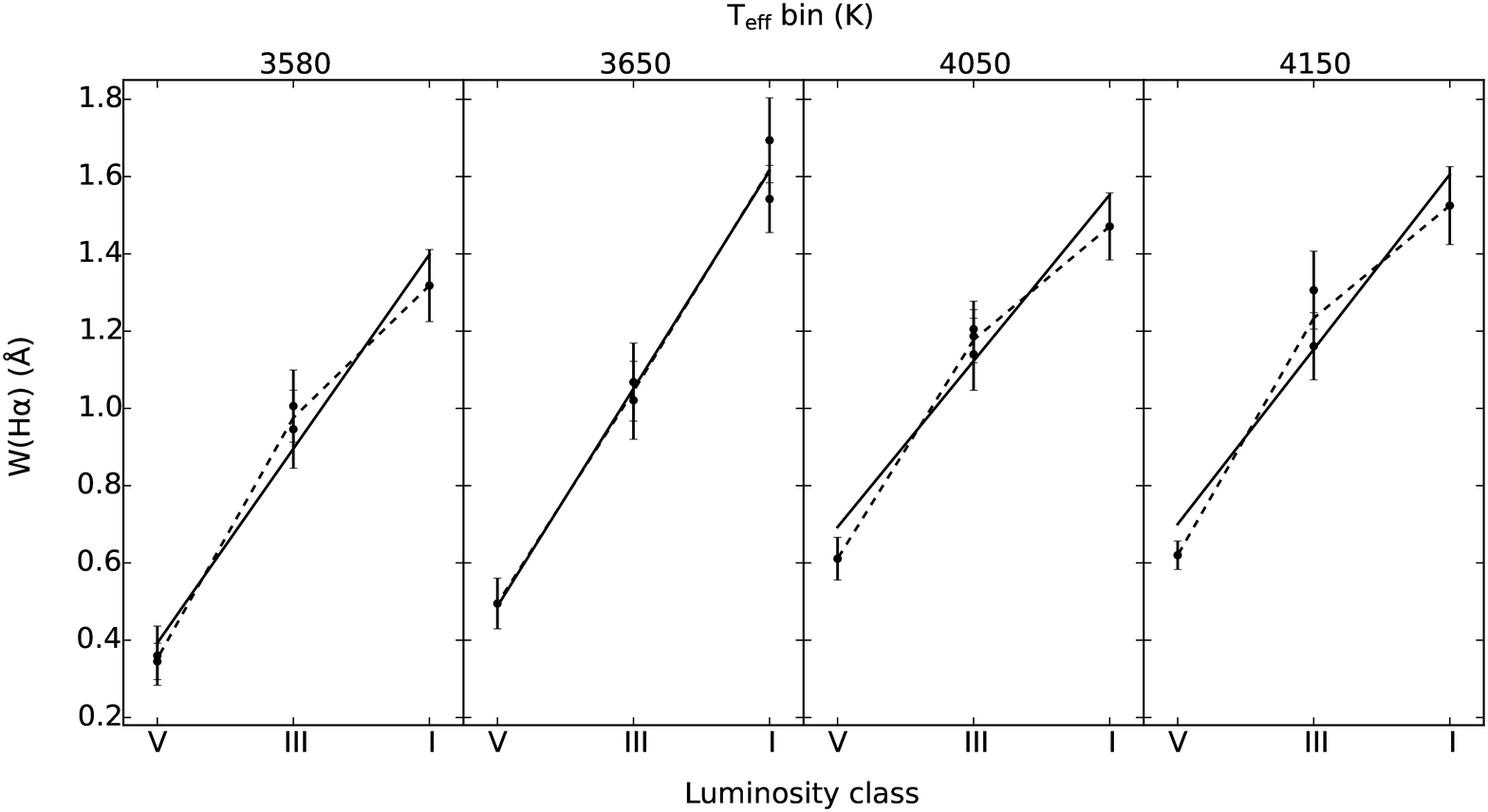}
\plotone{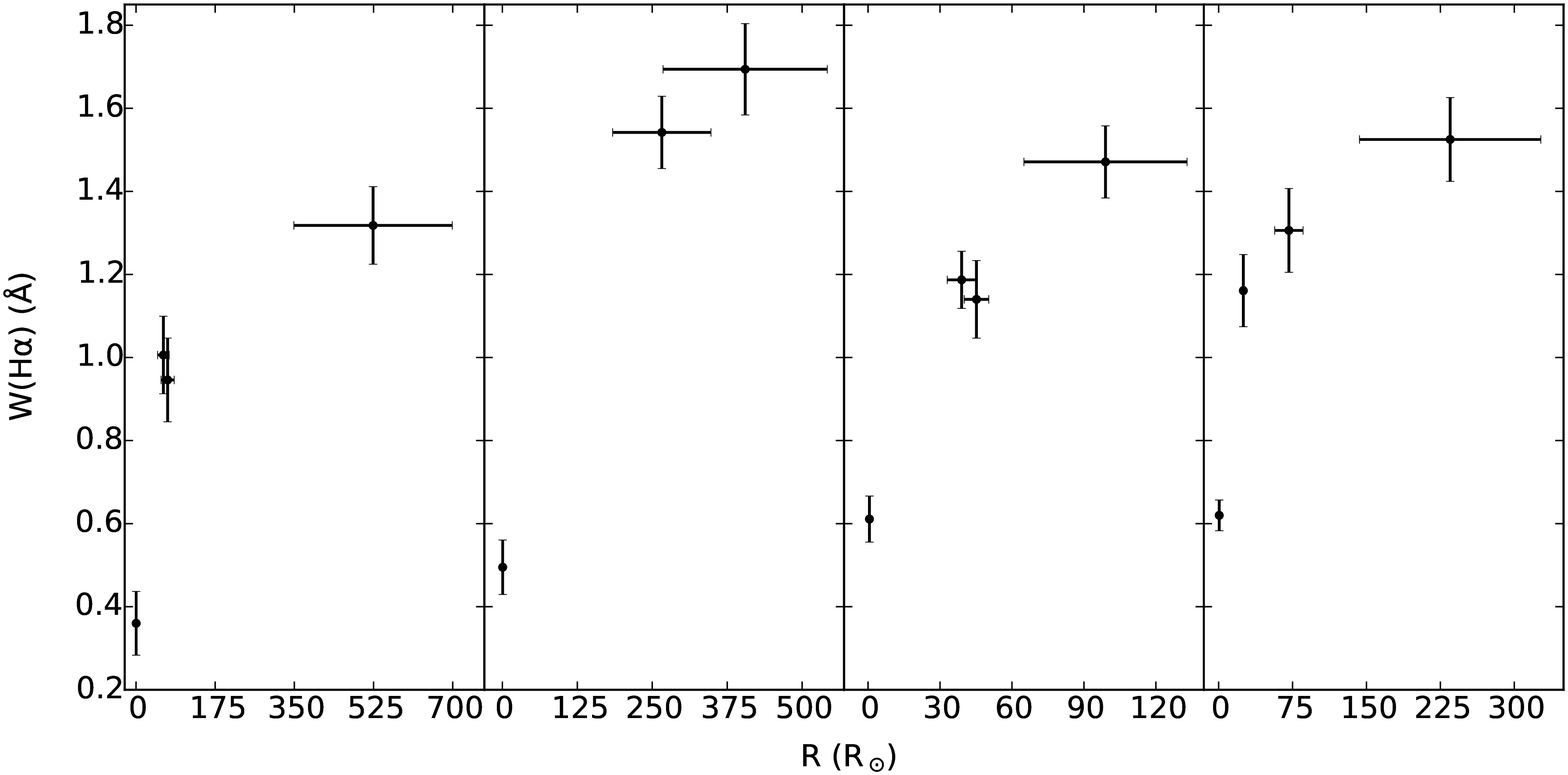}
\caption{\textit{W}(H$\alpha$), the H$\alpha$ equivalent width, for each star in our sample versus luminosity class (top) and stellar radius where available (bottom). Solid and dashed lines are as in Fig.2. Trends are consistent with \textit{W}(CaT), while the mean fractional difference in \textit{W}(H$\alpha$) is $-$1.00 between luminosity classes I and V, $-$0.30 between I and III, and $-$0.68 between III and V.}
\end{figure}

\begin{figure}
\epsscale{0.8}
\plotone{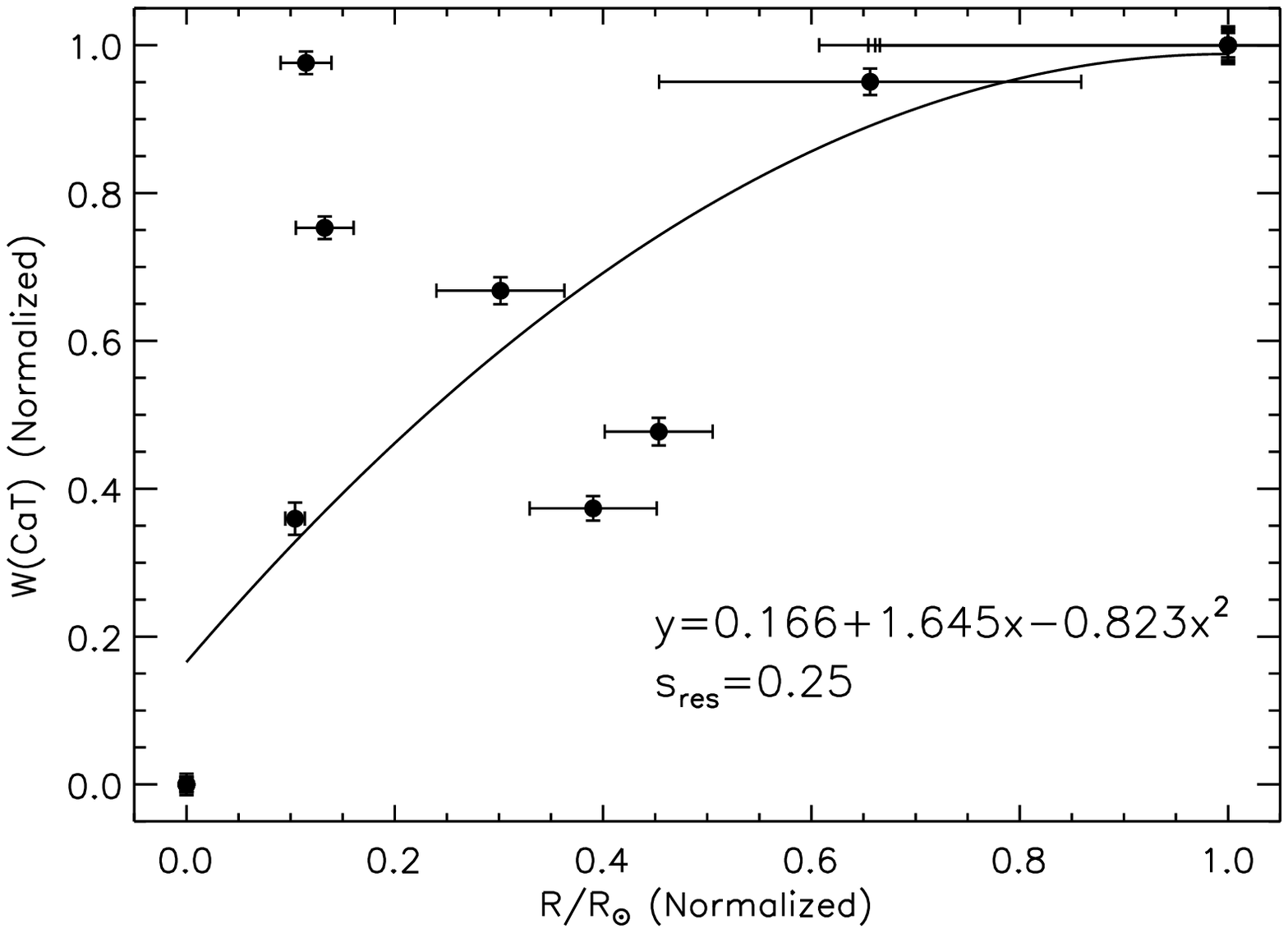}
\plotone{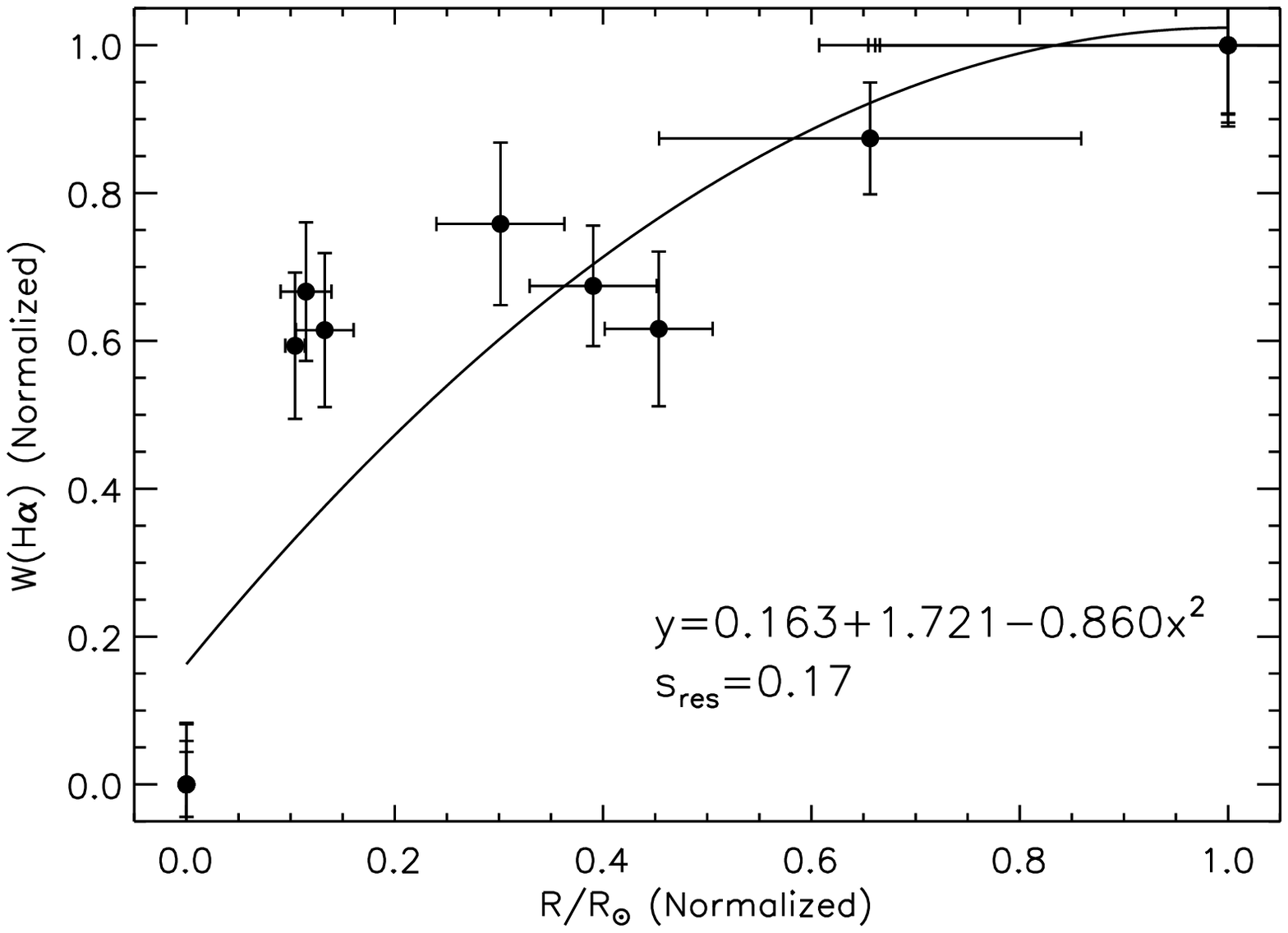}
\caption{A comparison of $W$(CaT) (top) and $W$(H$\alpha$) (bottom) to stellar radius for our full sample of stars. The data have been normalized within each of the four $T_{\rm eff}$ bins and then combined. For both luminosity class diagnostics the best-fit second-order polynomial is included and illustrated as a solid line along with the residual standard deviation ($s_{res}$).}
\end{figure}
\end{document}